\documentclass{cernyrep}

\usepackage[utf8x]{inputenc}   
\DeclareUnicodeCharacter{"2061}{}    
\usepackage{hyperref} 
\hypersetup{colorlinks=true, citecolor=black, linkcolor=black, urlcolor=blue} 

\usepackage{varwidth}
\usepackage{xcolor}
\newcommand\query[1]{}%
\usepackage{fancyhdr}
\fancyhfoffset{4 mm}
\fancypagestyle{ARTTITLE}{%
\fancyhf{} 
\chead{\small Proceedings of the 2017 CERN--Accelerator--School course on \it Vacuum for Particle Accelerators, Glumsl\"ov, (Sweden)} 
\lfoot{Available online at \url{https://cas.web.cern.ch/previous-schools}}
\rfoot{\thepage\hspace*{3mm}}
 
}

\begin{document}
\title{Brief Introduction to Particle Accelerators}
\author{P. F. Tavares}
\institute{MAX IV Laboratory, Lund, Sweden}

\begin{abstract}
This lecture is a brief introduction to charged particle accelerators. The aim is to provide the reader with basic concepts and tools needed to describe the motion of charged particles under the action of guiding and focussing fields, with an emphasis on those aspects that are relevant to understanding and quantifying how accelerator vacuum systems affect accelerator performance. Even though the focus is on electron accelerators and, in particular, electron storage rings used as synchrotron light sources, most of the concepts described are of general application to a wider class of particle accelerators.
\end{abstract}


\maketitle
\thispagestyle{ARTTITLE}

\query{Please check that all the figures make sense when reproduced in black and white (greyscale). Please check that you have obtained the necessary permissions to reproduce any copyright figures, and that this information is included in the appropriate figure captions.}

\query{Please check all figure labels carefully. Use correct symbols. Ensure that the maths in all figure labels is formatted correctly; variable in italic, labels in roman, \etc, and that text labels are consistently capitalized (\eg all lower case, initial letter upper case.}

\section{Assumptions and Goals}
In writing these notes, I have assumed that the reader is familiar with relativistic mechanics and electromagnetic theory at the advanced undergraduate or beginning graduate level. I have not assumed any previous knowledge of accelerators or charged particle beam dynamics. The aims of this lecture are:
\begin{enumerate}
	\item to provide motivations for developing and building  particle accelerators;
	\item to describe the basic building blocks of a typical particle accelerator;
	\item to describe the basic concepts and tools needed to understand how the vacuum system affects accelerator performance.
\end{enumerate}

Caveat: I will focus the discussion/examples mostly concerning one type of accelerator, namely electron storage rings optimized as sources of synchrotron radiation, although most of the discussion can be translated into other accelerator models.

The interested reader can find all of the material presented here discussed at length in a number of by now classical references.
The bibliography section lists some of them (\cite{Helmut},\cite{Sands},\cite{Edwards}), which have been particularly useful in preparing these notes.

\section{Why particle accelerators ?}
\begin{figure}[ht]
	\begin{center}
		\includegraphics[width=12cm]{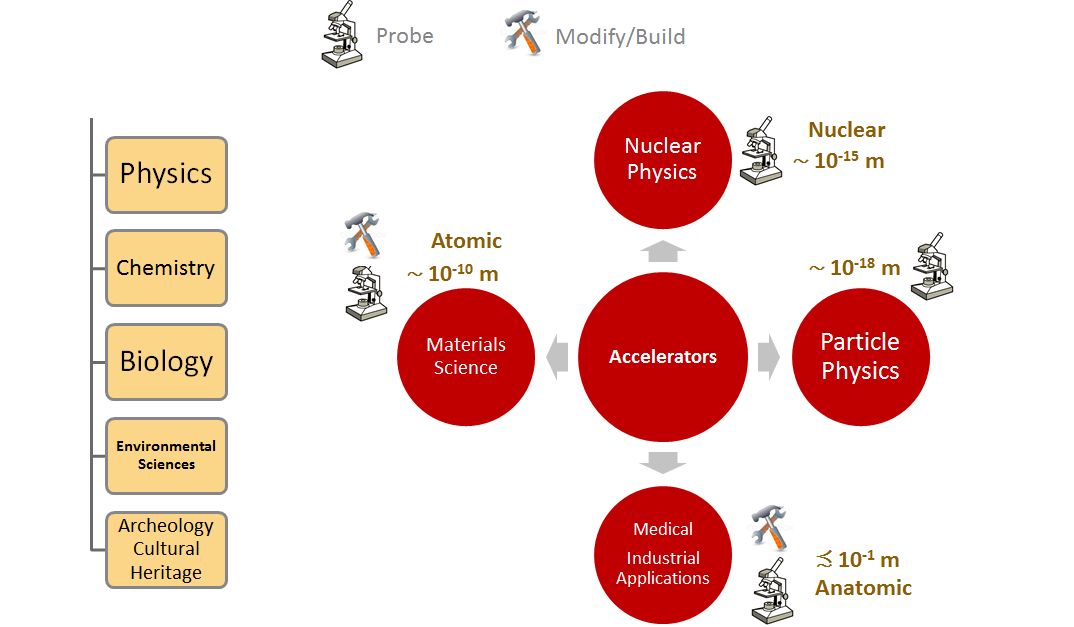}
		\caption{Particle accelerators and their applications either as probes into the properties of matter at various length scales or as tools to modify those properties and build new materials.}
		\label{fig:apppartacc}
	\end{center}
\end{figure}

Devices that produce and accelerate subatomic particles are essential instruments of investigation in various fields of basic and applied science. Although originally born in the 1930s out of the needs of nuclear physics to use high energy charged particle beams to probe into the inner workings of matter at the scale of nuclei, these machines have by now conquered much wider fields of application in science, medicine, and industry  to probe into and/or modify properties of matter at multiple length scales ranging from the sub-atomic (the realm of particle physics) to the anatomic level. In this lecture, I will focus on accelerators that produce beams that are useful in a wide range of fields in basic and applied science, which I designate here by the general name of {\em materials science} (Fig. \ref{fig:apppartacc}).

Several types of beams can be used as probes into matter -- photons (electromagnetic radiation), neutrons, as well as ions -- and particle accelerators can be used to produce all of these beams. Let me focus on photons and neutrons: these are complementary techniques which each have specific characteristics which make them more suitable for certain types of problems,  while photons will interact mostly with the atomic electrons in the samples, neutrons will interact with the nuclei. Neutrons will be more penetrating and have stronger interactions with materials of low atomic number (such as hydrogen). Another feature is that neutrons have an intrinsic magnetic moment which makes them suitable for probing into magnetic properties of materials. This is also possible with photons but typically one needs to look at resonant phenomena for which the interaction cross-sections grow significantly.  Finally one very important difference is that photon beams are typically many orders of magnitude brighter (brightness being defined as the particle flux per unit source area and unit divergence) than neutron beams, which makes photons unbeatable, for example, in applications where one needs to focus the beam onto very small spots.

Photon sources can be further classified as based on circular accelerators (storage rings) and free electron lasers based on linear accelerators. Here again, the intended applications will define the best choice of instrument: storage-ring-based sources, which I will focus on for the rest of this lecture, excel in average photon flux and average brightness whereas free-electron lasers will achieve much higher peak brightness in very short light pulses.

An example of an accelerator-based neutron source is the European Spallation Source\cite{ess} currently under construction in Lund, whereas there are tens of examples of storage-ring-based synchrotron light sources worldwide, with the MAX IV facility\cite{maxivjsr} in Lund being the most recent addition to the club.

\subsection{Why synchrotron light sources ?}
Synchrotron radiation or synchrotron light is electromagnetic radiation emitted by charged particles undergoing acceleration.  One specific type of acceleration, namely centripetal acceleration, occurs  which the acceleration vector is perpendicular to the velocity vector, as when a charged particle has its path deflected by the action of a magnetic field. At low (non-relativistic) velocities the emitted radiation exhibits a wide angular aperture, but as the particle energy increases and reaches the ultrarelativistic limit (i.e., for particle velocities very close to the velocity of light), the angular distribution of the emitted radiation changes dramatically and an extremely narrow cone of very wideband radiation emanates from the charged particle beam along the instantaneous direction of motion of the particle. This collimation effect is a feature common to all systems in which waves are emitted by a source moving at a speed close to the speed of the waves\cite{Margaritondo95} and becomes extreme in the case of synchrotron radiation emitted by high energy particles. In fact, the most interesting properties of synchrotron radiation are a consequence of this very basic geometrical property resulting from the ultrarelativistic speeds.

The modern builders and users of synchrotron light sources are, in a sense, following the steps of others who used light as a probe to obtain information on the natural world. Two obvious examples extending from the observation of very large and distant objects to the observation of extremely small objects come immediately to mind: at the extreme of very large objects we may think of Galileo who used optical instruments (telescopes) to change our whole picture of the universe. In the opposite extreme of the very small, we can think of the Dutch cloth merchant Anton von Leeuwenhoek, who was the first to use an optical microscopes (or rather a magnifying glass) to observe bacteria. Von Leeuwenhoek observed, for example a bacterium named Selenomonas commonly found in human mouths. It is a pretty big bacterium, about ten micrometres long, so it was possible for Leeuwenhoek's microscopes, with magnifications of a few hundred times, to see them. Even with modern optical microscopes, resolutions are typically limited to some several hundred nanometres due to diffraction, i.e., images of an object illuminated with light at a given wavelength will become blurred when object dimensions become close to the wavelength of the light.

To improve on that and to be able to probe into much smaller structures, you only need to realize that visible light corresponds to but a small fraction of the electromagnetic spectrum, so that if you can find a source of radiation at shorter wavelengths than visible light, say UV light or X-rays, than you can probe into structures with sizes comparable to molecules and atoms. That is exactly where synchrotron radiation comes into the picture.

With synchrotron radiation, a wide variety of materials can be studied at the atomic and molecular level. One can, for example, determine the geometry, i.e., the spatial arrangement of the many atoms that make up macromolecules such as proteins, that plays an important role in the physiology of living organisms. This is done by a technique known as protein crystallography in which one prepares crystals by arranging these molecules in a regular pattern and then shining  X-rays onto them and observing the resulting diffraction patterns. Particularly relevant for a number of applications is the fact that synchrotron radiation is tuneable, i.e., the radiation wavelength can be chosen over a wide range with great accuracy allowing a number of spectroscopic methods to be implemented. Additionally, synchrotron radiation has definite polarization properties that can be put to use for example in the study of the magnetic properties of matter. Moreover, synchrotron light comes in the form of light pulses and this temporal structure can be made use of to perform time-resolved studies. Finally, in recent years the large increase in the radiation brightness and corresponding increase in coherence have enabled a number of new imaging methods to be developed.

\begin{figure}[ht]
	\begin{center}
		\includegraphics[width=12cm]{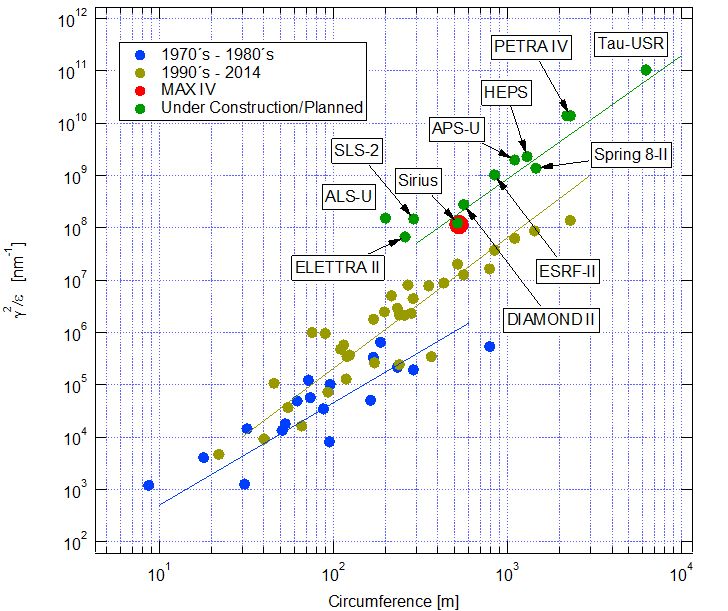}
		\caption{Evolution of storage-ring-based synchrotron light source performance. The horizontal axis is the storage ring circumference and the vertical axis is the inverse storage ring emittance normalized to the electron beam energy -- a quantity that under certain conditions is proportional to the photon beam brightness. Each dot in this diagram corresponds to a source and the colours indicate the year of commissioning. The green dots corresponds to proposed  or planned future sources.}
		\label{fig:brightevol}
	\end{center}
\end{figure}

Storage-ring-based light sources  are typically classified according to the {\em generation} they belong to. The first generation (built in the seventies) comprises machines originally designed for nuclear physics applications and  were used parasitically as light sources. The second generation sources (built in the eighties) were fully dedicated to synchrotron light from bending magnets, whereas the third generation sources (built in the 1990s and 2000s) were optimized for insertion devices\footnote{In modern storage rings, most of the radiation actually used for materials research comes not from the magnets that cause the beam to go around in a circle, but rather from special magnets called undulators and wigglers. In these devices, a periodic array of magnet blocks produces an approximately sinusoidal variation of the deflecting field along the beam bath causing the electron beam to {\em undulate} sideways, allowing the production of much more light by combining radiation coming from different periods.} and aimed at high brightness. In the past few years, a new generation, sometimes referred to as (near) diffraction-limited light sources, has appeared and the first example of this new generation is the MAX IV 3 GeV ring, recently commissioned in Lund, Sweden. Several projects worldwide are today (Fig. \ref{fig:brightevol}) following the same trend, resulting in an impressive increase of many orders of magnitude in source brightness.

\section{The electron storage ring: {\em accelerator building blocks}}

In order to generate synchrotron radiation, a beam of high energy particles (typically electrons or po\-si\-trons) needs to be generated and then made to go around on an approximately circular path over many turns, emitting vast amounts of radiation every time its trajectory is bent. High energy particle are needed because the spectrum and amount of radiation depends strongly on the particle energy. In fact, it takes typically electrons of a few GeV to generate synchrotron light photons in the few keV energy range. The reason for using electrons or positrons instead of heavier particles, such as protons, is that the amount (power) of the emitted radiation is inversely proportional to the fourth power of the particle mass. Electrons are often the preferred choice as they are much cheaper to produce, although some early facilities did use positrons  to avoid problems related to the trapping of positively charged ions by the beam.

\begin{figure}[ht]
	\begin{center}
		\includegraphics[width=12cm]{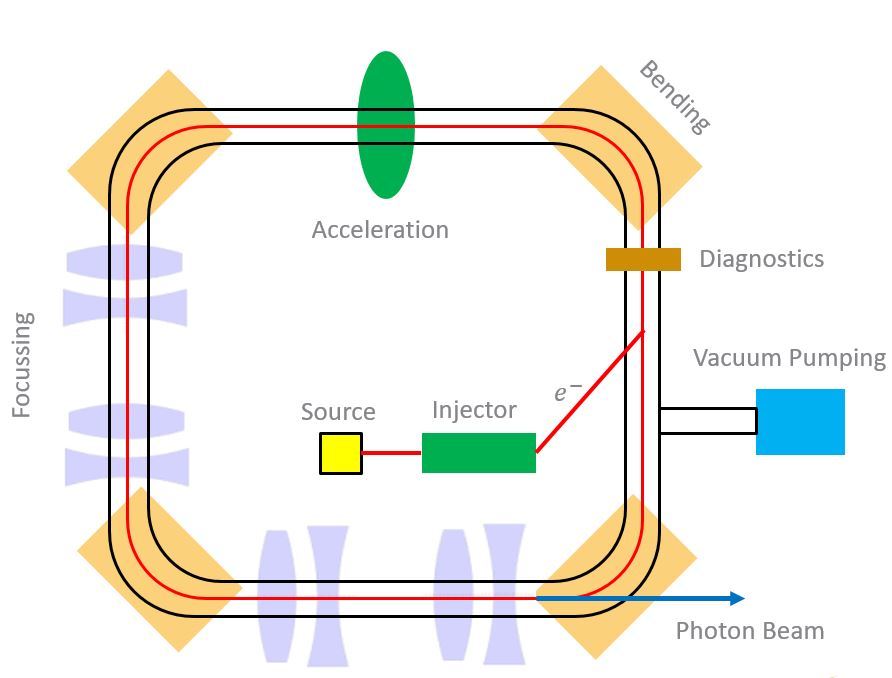}
		\caption{Schematic view of an electron storage ring}
		\label{fig:schematic}
	\end{center}
\end{figure}

Figure~\ref{fig:schematic} illustrates the building blocks of a typical storage ring. The electron beam is produced at a source, pre-accelerated by an injector system, and then transferred to the storage ring. The electron sources are called electrostatic or radio-frequency electron guns, depending on whether the acceleration in them is achieved by static electric fields or high frequency rf (radio frequency) fields. \query{Please define rf.} In both cases, electrons emanate from metallic cathodes either by thermionic emission or by incidence of a laser pulse on the cathode surface. The injector systems can be of various types: a linear accelerator followed by an intermediate booster synchrotron ring is a common choice at many labs, but the use of a linear accelerator as a full energy injector is also possible. In most cases, the amount of charge produced by the injector system at each pulse is not enough to produce the desired high circulating currents in the storage ring and a sequence of injector pulses need to be accumulated or stored in the ring.

In order to guide and focus the high energy beam along a closed path, static magnetic fields are typically used. In fact, even though electrostatic fields could in principle be used, magnetic fields are preferred at high energies due to technological limitations. DC electric fields are typically limited to about $10^{7}$ V/m due to voltage breakdown (sparking), whereas magnetic fields up to about 2 T (room temperature magnets) or 5--10 T (superconducting magnets) can be routinely obtained. As a result, the maximum bending force that can be obtained using magnetic fields is about two orders of magnitude larger that the corresponding maximum electrical forces for ultrarelativistic particles.

Moreover, it is necessary to replace the energy lost by the  charged particles as they emit synchrotron radiation, which is done at metallic resonating cavities in which stationary oscillating electric and magnetic fields are established, with a longitudinal component of the electric field giving energy to the beam at each passage of the beam through the cavity. In addition, a vacuum system is needed in order to guarantee that the path along which particles travel are sufficiently free of gas molecules preventing collisions that could lead to beam particle losses or deterioration of particle beam quality. Finally, a number of diagnostic  tools are needed to characterize beam parameters (charge, current, position, size, lifetime) and in certain cases to act back on the beam to guarantee stable and reliable operation.

\section{Basic beam dynamics in storage rings}.
\label{sect:bdsr}
The first task that the electron storage ring designer is confronted with is to define spatial and temporal configurations of external\footnote{By external, I mean we do not consider fields generated by the electron beam itself.} electromagnetic fields that can:

\begin{enumerate}
	\item guide the beam along a design orbit;
	\item maintain  the momentum/energy of the electrons despite the energy loss to synchrotron radiation;
	\item keep electrons whose position, angle, or momentum slightly deviates from the design values in the vicinity of the design orbit and of the design energy.
\end{enumerate}

\begin{figure}[ht]
 	\begin{center}
 		\includegraphics[width=6cm]{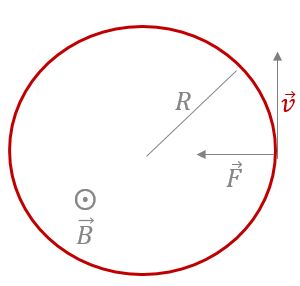}
 		\caption{Design orbit in a circular accelerator with homogeneous (azimuthally symmetric) bending field}
 		\label{fig:cyclotron1}
 	\end{center}
 \end{figure}

\begin{figure}[ht]
 	\begin{center}
 		\includegraphics[width=6cm]{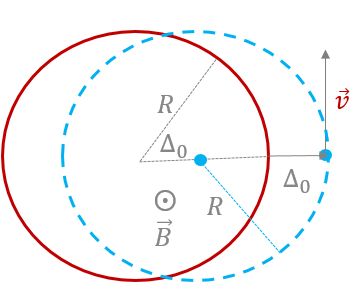}
 		\caption{The orbit of an electron that has a small deviation in position with respect to the nominal orbit, but has otherwise the nominal momentum and angle.}
 		\label{fig:cyclotronperturbed}
 	\end{center}
 \end{figure}

Whereas the first two points above can be approached with the straightforward use of the Lorenz force equation and Newton's equation of motion:

\begin{equation}
\vec{F}=q\left(\vec{E}+\vec{v}\times{}\vec{B}\right)=\frac{d\vec{p}}{dt} \, ,
\end{equation}

the third point is in fact the one that requires most ingenuity and detailed analysis. A real beam is after all composed of a large number of electrons (a 500 mA circulating beam in the MAX IV 3 GeV storage ring, for example, contains  $\approx 5 \times 10^{12}$ electrons)  each having position, angle, and energy that slightly deviates from the nominal values and it is easy to see that special care needs to be taken to guarantee that such initially small deviations do not grow without bounds and eventually lead to particle loss. Take for example the simple case of a perfectly homogeneous magnetic guide field pointing in the vertical direction $z$ as indicated in Fig.~\ref{fig:cyclotron1}. The design orbit for particles with nominal momentum $p$ is simply a circle with radius $R$ determined by the magnitude $B$ of the magnetic field
\[ R=\frac{p}{e_0B} \] where $e_0$ is the electronic charge. In order to analyse motion stability, let us  consider the motion of an electron that has the nominal momentum and angle but whose initial position differs from the nominal position by $\Delta_0 \ll R$ (Figure~\ref{fig:cyclotronperturbed}). Since the field is assumed homogeneous, the trajectory of this deviating electron is clearly also a circle of same radius, but whose centre is shifted by the same amount $\Delta_0$. Using polar coordinates $(\rho ,\theta)$ in a reference system with its origin at the centre of the nominal orbit, we can write the equation for the motion of the perturbed electron as:

 \begin{equation}
 {\rho{}(t)}^2  = {\left({\Delta{}}_0+R\cos{\omega{}t}\right)}^2+{\left(R\sin{\omega{}t}\right)}^2 \, ,
  \end{equation}

where $\omega = \frac{v}{R}=\frac{e_0vB}{p}$ is the angular cyclotron frequency. The time evolution of the deviation of the actual orbit with respect to the nominal orbit ${\rho{}}_0(t)=R$
can then be written as

\begin{equation}
\Delta{}\left(t\right)=\rho{}\left(t\right)-{\rho{}}_0\left(t\right)=\rho{}\left(t\right)-R=\frac{2{\Delta{}}_0R\cos{\omega{}t}+{\Delta{}}_0^2}{\rho{}\left(t\right)+R}\, ,
\end{equation}

and, by Taylor expanding and dropping all terms of order higher than one in $\frac{\Delta_0}{R}$, we obtain:

\[
\Delta{}(t)\approx{}{\Delta{}}_0\cos{\omega{}t}\, .
\]
 In other words, an initially small deviation $\Delta_0$ {\em remains small} and the perturbed electron simply oscillates stably around the nominal electron. The simple homogeneous magnetic field configuration does provide conditions for stable motion over many turns on the plane of the orbit. A similar conclusion can be reached by analysing an electron that starts out its motion with the nominal momentum and position, but which has its velocity pointing in the wrong direction, i.e., has a small angular deviation.

\begin{figure}[ht]
 	\begin{center}
 		\includegraphics[width=5cm]{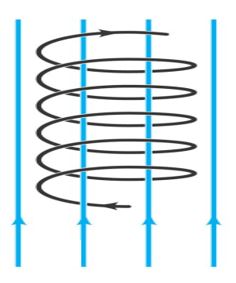}
 		\caption{Orbit of an electron that has a small component of its initial velocity parallel to the magnetic field}
 		\label{fig:spiral}
 	\end{center}
 \end{figure}

If, on the other hand, the perturbed electron has a small velocity component along the direction perpendicular to the plane of the orbit (i.e., the direction parallel to the direction of the bending field), the situation is different (Fig.~\ref{fig:spiral}). Indeed, we can see from the Lorenz force equation above that the motion along that direction will become a uniform linear motion since there is no magnetic force and the final motion is a spiral that leads particles to acquire larger and larger deviations from the nominal orbit along the vertical plane. In accelerator jargon, we would say that the homogeneous field configuration lacks vertical focussing and such an accelerator does not allow a beam to be stably stored.

\begin{figure}[ht]
	\begin{center}
		\includegraphics[width=14cm]{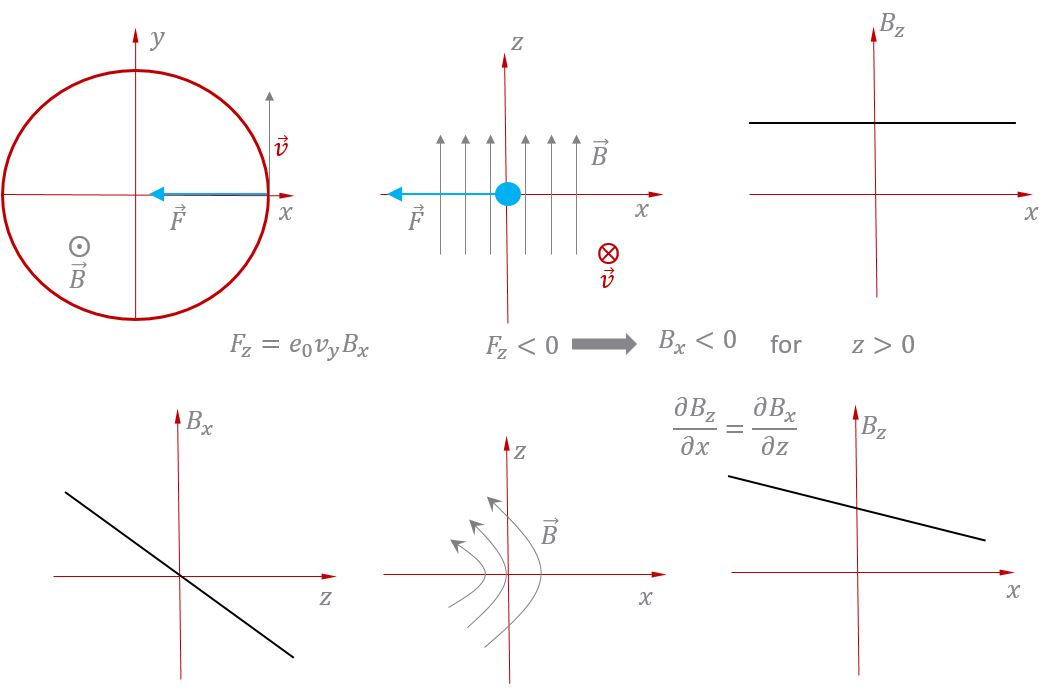}
		\caption{How to obtain vertical focussing in an azimuthally symmetric circular accelerator}
		\label{fig:VerticalFocussing}
	\end{center}
\end{figure}

In order to provide vertical focussing, a horizontal component of the magnetic field is required so that a vertical restoring force will kick the beam back towards the plane of the nominal orbit whenever the electrons are either above or below that plane. As illustrated in Fig.~\ref{fig:VerticalFocussing}, this requires a negative horizontal field component above the plane of the nominal orbit and a positive horizontal field component below the nominal design orbit, in other words, the horizontal field component needs to have a negative gradient as a function of radial position, i.e.,
\begin{equation}
\frac{\partial{}B_\mathrm{x}}{\partial \mathrm{z}} < 0 \, , \label{eq:vertstab}
\end{equation}

on the plane of the orbit. Since Maxwell's equations require that

\begin{equation}
\frac{\partial{}B_\mathrm{z}}{\partial{}\mathrm{x}}=\frac{\partial{}B_\mathrm{x}}{\partial{}\mathrm{z}} \, ,
\end{equation}

the vertical component of the bending field also has to vary along the radial direction. In fact, the absolute value of the vertical component needs to decrease along the radial direction in order to guarantee vertical stability. Such a spatial field distribution can be obtained by properly shaping the pole pieces of the bending magnet, namely by making the pole gap increase as a function the distance from the centre of the nominal orbit.

\begin{figure}[ht]
 	\begin{center}
 		\includegraphics[width=10cm]{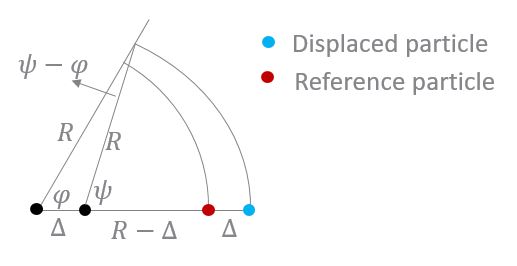}
 		\caption{Horizontal focussing in a homogeneous field}
 		\label{fig:Slice}
 	\end{center}
 \end{figure}

Even though the field distribution described above does produce stable motion in both the horizontal and vertical planes (and a number of accelerators were built based on this {\em weak focussing} principle in the early days), it suffers from a fundamental limitation: the strength of the focussing is severely limited implying the need for large magnet apertures. In fact, in order to increase the strength of the focussing in the vertical direction, a large (negative) gradient of the vertical field component would be needed, but this at the same time reduces the strength of the focussing in the horizontal direction. To understand this, let us go back to the physical origin of radial focussing in a homogeneous field. Take a slice of bending field that bends the nominal electron by an angle $\varphi$  (Fig.~\ref{fig:Slice}). An electron displaced to the outside of the nominal orbit is bent by a larger amount than the nominal electron, since it sees a longer path length under the action of the homogeneous bending field as it goes through the same slice. The larger bending can be described as a an additional focussing kick that is seen by the electron displaced to the outside of the nominal orbit and, given by:

\begin{equation}
\delta{} = \psi -\varphi \approx{}\frac{\Delta{}}{R}\sin{\varphi{}}\, ,
\end{equation}

where again we assume $\Delta \ll R$. This is in fact the physical origin of the radial focussing properties of a homogeneous field which we derived earlier. If now the vertical bending field decreases radially, it is clear that at some value of field gradient the increased path  length effect will be exactly cancelled by the fact that field is smaller at larger distances from the centre of the nominal orbit and the horizontal focussing will vanish. This happens indeed when

 \begin{equation}
 n=\frac{R}{B_\mathrm{z}}\frac{\partial{}B_\mathrm{z}}{\partial{}x}<-1 \, ,\label{eq:fieldindex}
 \end{equation}

 where $n$ is called the {\em field index}. Combining this result with the vertical stability condition Eq.~(\ref{eq:vertstab}) we see that stability on both planes requires

 \begin{equation}
-1<n<0 \, . \label{eq:estabwf}
 \end{equation}

\begin{figure}[ht]
	\begin{center}
		\includegraphics[width=10cm]{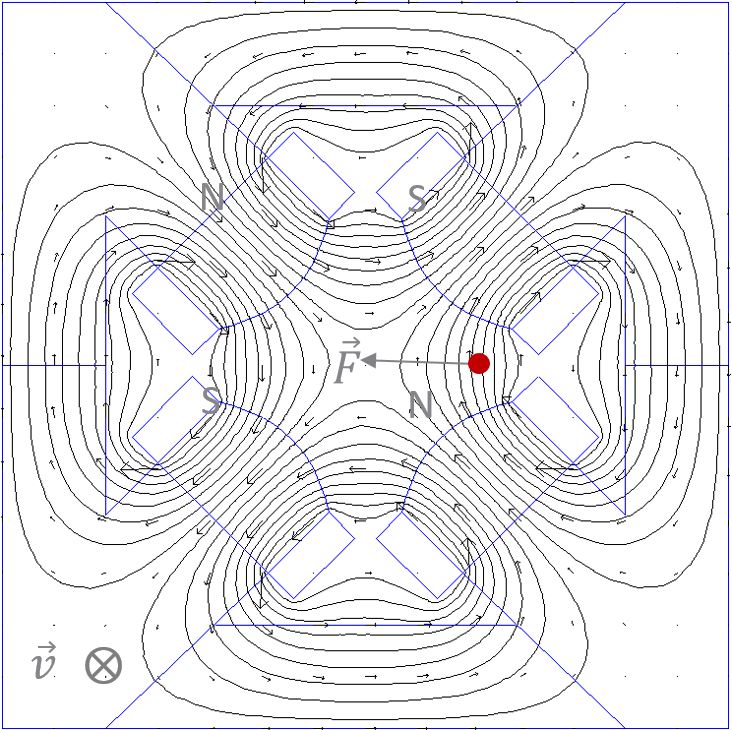}
		\caption{A horizontally focussing quadrupole. An electron going into the plane of the figure that is shifted to the right is bent back towards the centre.}
		\label{fig:quadexemple}
	\end{center}
\end{figure}

\begin{figure}[ht]
	\begin{center}
		\includegraphics[width=12cm]{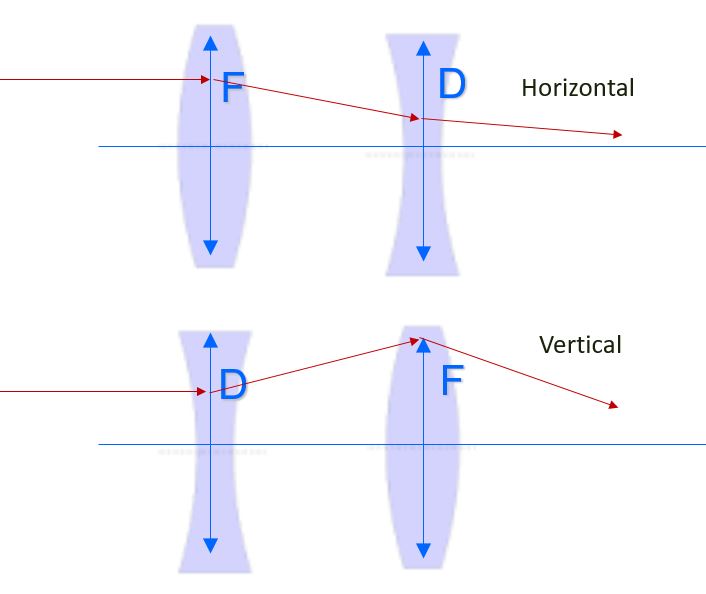}
		\caption{How a sequence of focussing and defocussing elements can be used to achieve an overall focussing effect}
		\label{fig:strongfocus}
	\end{center}
\end{figure}

 The solution of the dilemma imposed by the weak focussing limits above was the introduction in the 1950s of the  alternating gradient or strong focussing concept. In fact, if one breaks the assumption of azimuthal symmetry implicit in our previous considerations and considers a sequence of segments in which either only horizontal or only vertical focussing is provided, it is possible to devise a configuration in which stability for both planes is achieved while allowing focussing strengths to grow. This requires that an appropriate ratio of the distance between focussing/defocussing elements to their integrated focussing strengths is maintained. Figure~\ref{fig:quadexemple} illustrates how a quadrupole magnet may be used to provide a section where particles are focussed in the horizontal plane and simultaneously defocussed in the vertical plane whereas Fig.~\ref{fig:strongfocus} shows how an overall focussing effect can be obtained in both planes by a proper combination of focussing strengths and distances between the quadrupoles. In the horizontal plane, particles are first focussed and then defocussed, but since they cross the defocussing region at lower amplitudes, the defocussing kick is smaller than the focussing one and an overall focussing effect is achieved. In the vertical plane, particles cross first a defocussing region and go subsequently through the focussing region at larger amplitudes again where they get a larger focussing kick, again leading to an overall focussing effect.

\subsection{Transverse beam dynamics}

\begin{figure}[ht]
	\begin{center}
		\includegraphics[width=12cm]{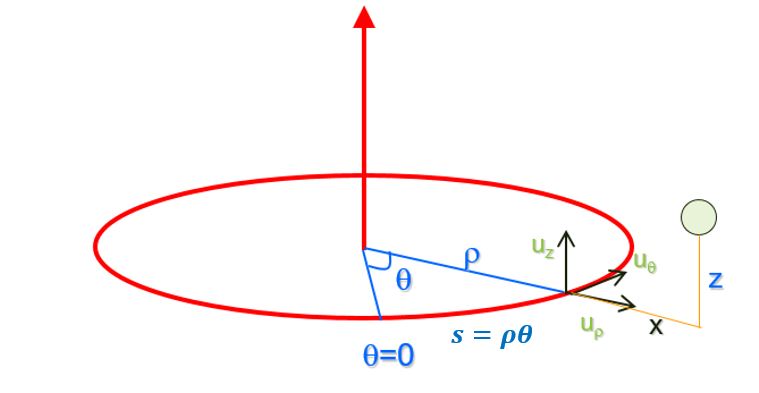}
		\caption{Coordinate system definition}
		\label{fig:CoordinateSystem}
	\end{center}
\end{figure}

The rather cursory discussion above can be put onto solid mathematical grounds and the interested reader is referred to the references at the end of this note for a detailed treatment. Here, I only mention a collection of results that are useful in describing accelerator performance parameters and quantifying the effects of vacuum system on the dynamics of the beam.
A general analysis of beam motion in a storage ring starts from the establishment of a  coordinate system $(s,x,z)$ in which small deviations with respect to the nominal orbit can be conveniently expressed (Fig. ~\ref{fig:CoordinateSystem}).

The magnetic guide field is typically assumed to obey the symmetry conditions

\begin{equation}
B_\mathrm{x}\left(s,x,z\right)={-B}_\mathrm{x}\left(s,x,-z\right)
\end{equation}

\begin{equation}
B_\mathrm{z}\left(s,x,z\right)=B_\mathrm{z}\left(s,x,-z\right)\, ,
\end{equation}

and can be written, to first order in deviation from the design orbit, as:
\begin{equation}
B_\mathrm{z}\left(s,x,z\right)=B_0\left(s\right)+\frac{\partial{}B_\mathrm{z}}{\partial{}x}x=B_0\left(s\right)-g(s)x
\end{equation}

\begin{equation}
B_\mathrm{x}\left(s,x,z\right)=\frac{\partial{}B_\mathrm{x}}{\partial{}z}z=-g(s)z
\end{equation}

\begin{equation}
B_{\theta{}}=0 \, ,
\end{equation}

where the transverse field gradient is given by

\begin{equation}
g(s)=-\frac{\partial{}B_\mathrm{z}}{\partial{}x}=-\frac{\partial{}B_\mathrm{x}}{\partial{}z}\, .
\end{equation}

Note that the bending field is purely vertical on the plane of the orbit and that only transverse components are considered, i.e., edge effects at the ends of magnets are disregarded.

The Lorentz force equation then leads to the following equations of motion for the horizontal ($x$) and vertical ($z$) deviations with  respect to the nominal orbit for a particle of momentum $p=p_0+\Delta p$, where $p_0$ is the nominal momentum and $\Delta p \ll p_0$, is:

\begin{eqnarray}
x^{''}(s) + \left[ \frac{1}{\rho (s)^2} - K(s) \right] x(s) & = & \frac{1}{\rho (s)} \frac{\Delta p}{p_0} \label{eq:eqmotion1} \\
z^{''}(s) + K(s)z(s) & = & 0 \, ,
\label{eq:eqmotion2}
\end{eqnarray}

where the {\em paraxial approximation} conditions
\begin{eqnarray}
x & \ll & \rho \\
x^{'}\left(s\right) & = & \frac{dx}{ds}\ll{}1 \\
z^{'}\left(s\right) & = & \frac{dz}{ds}\ll{}1
\end{eqnarray}

have been assumed to hold. The local bending radius and normalized transverse field gradient are given by

\begin{eqnarray}
K(s) & =& \frac{e_0 g(s)}{p_0} \\
\frac{1}{\rho(s)} & = & \frac{e_0 B_0(s)}{p_0} \, ,
\end{eqnarray}

and the path length $s$ along the nominal orbit is used as the independent variable instead of time (primes denote derivatives with respect to $s$). For an {\em on-energy} particle ($\Delta_\mathrm{p} = 0$), Eqs.~(\ref{eq:eqmotion1}) and ~(\ref{eq:eqmotion2}) can both be written as
\begin{equation}
u^{''} + K_\mathrm{u}(s)u(s) = 0  \label{eq:hill}
\end{equation}
where $u$ is either $x$ or $z$ and
\begin{eqnarray}
K_\mathrm{x} (s) & = & \frac{1}{\rho (s)^2} - K(s) \\
K_\mathrm{z} (s) & = & K(s)\, .
\end{eqnarray}

Note that, since we treat a circular accelerator, the functions  $\rho (s)$ and $K(s)$ are clearly periodic with a period equal to the accelerator circumference. Actually, in most practical cases, an even stricter periodicity condition is satisfied as storage rings are often composed of smaller units (superperiods) that repeat themselves along the circumference.
With the equations of motion cast into this form, one can express the question of stability of motion in a circular accelerator as that of finding conditions for the distribution of external fields given by the functions $\rho (s)$ and $K(s)$ that lead to oscillatory solutions. In accelerator jargon this is often referred to as designing the accelerator's magnet lattice. Equation~(\ref{eq:hill}) with the periodicity conditions above are known as Hill's equations and were first introduced in the late 19th century to study the motion of the moon.

A general solution to Eq.~(\ref{eq:hill}) can be written in the {\em pseudo-harmonic} form:
\begin{equation}
u(s)  =  \sqrt{\epsilon _\mathrm{u} \beta_\mathrm{u}(s)} \cos \left( \phi_\mathrm{u}(s) + \phi_0 \right) \, , \\
\label{eq:pseudosol}
\end{equation}
where $\epsilon_\mathrm{u}$ and $\phi_o$ are constants of the motion to be defined by the initial conditions and the betatron function $\beta_\mathrm{u}(s)$ and betatron phase $\phi_\mathrm{u}(s)$ satisfy
\begin{equation}
\phi_\mathrm{u}(s)  =  \int_{0}^{s} \frac{ds^{*}}{\beta_\mathrm{u}(s^{*})}
\end{equation}
\begin{equation}
\frac{1}{2}{\beta{}}_\mathrm{u}\left(s\right){\beta{}}_\mathrm{u}^{''}\left(s\right)-\frac{1}{4}{{\beta{}}_\mathrm{u}^{'}}^2\left(s\right)+{\beta{}}_\mathrm{u}^2\left(s\right)K_\mathrm{u}\left(s\right)  =  1 \, .
\label{eq:phase}
\end{equation}

Note that $\beta_\mathrm{u}$ is periodic with the same period as $K_\mathrm{u}(s)$. Particularly important is the phase advance calculated over a complete turn $C_0$ around the ring, related to the betatron tune:
\begin{equation}
Q_\mathrm{u} = \frac{1}{2\pi} \phi_\mathrm{u} \left( C_0 \right)\, .
\end{equation}
With the normalization factor in the definition above, $Q_\mathrm{u}$ becomes the number of oscillation periods per revolution.

Armed with these tools, we can now look again at the case of the weak-focussing accelerator that was described in section~\ref{sect:bdsr}. In an azimuthally symmetric ring, the bending and focussing strengths $\rho (s)=\rho_0$ and $K_{z}(s)=K_0$ , $K_\mathrm{x}(s)=\frac{1}{\rho_0^2} - K_0$ are independent of $s$ and Eq.~(\ref{eq:hill}) become simple harmonic oscillator equations with solutions:

\begin{eqnarray}
u \left( s \right)      & =  &  u_0 \cos \left( \sqrt{K_\mathrm{u}} s \right) + \frac{u_0^{'}}{\sqrt{K_\mathrm{u}}} \sin \left( \sqrt{K_\mathrm{u}} s \right) \label{eq:wkfctr2} \\
u^{'} \left( s \right)  & =  & -u_0 \sqrt{K_\mathrm{u}} \sin \left( \sqrt{K_\mathrm{u}} s \right) + u_0^{'} \cos\left( \sqrt{K_\mathrm{u}} s\right)\, ,
\label{eq:wkfctr}
\end{eqnarray}

where $u_0,u_0^{'}$ are the initial conditions. The stability condition for both planes, leading to bounded oscillation amplitudes, instead of exponentially growing motion reads then:

\begin{eqnarray}
K_\mathrm{z} & = & K_0 > 0  \\
K_\mathrm{x} & = & \frac{1}{\rho_0^{2}}-K_0 > 0 \, ,
\end{eqnarray}

which are equivalent to the conditions (Eq.~(\ref{eq:estabwf})) obtained in the previous section for the field index $n=\frac{{\rho{}}_0}{B_0}\frac{\partial{}B_\mathrm{z}}{\partial{}x}$. Inspection of Eqs.~(\ref{eq:wkfctr}) and (\ref{eq:pseudosol}) reveals furthermore that the betatron functions in this case are independent of $s$ and given by

\begin{eqnarray}
\beta_\mathrm{x} = \frac{1}{\sqrt{K_\mathrm{x}}} & = & \frac{\rho_0}{\sqrt{1-\rho_0^2 K_0}} \\
\beta_\mathrm{z} = \frac{1}{\sqrt{K_\mathrm{z}}} & = & \frac{1}{\sqrt{K_0}} \, ,
\end{eqnarray}

and that the initial conditions $\epsilon_\mathrm{u}$, $\phi_0$ in the pseudo-harmonic form are related to $u_0,u_0^{'}$ by
\begin{eqnarray}
u_0 & = & \sqrt{\epsilon_\mathrm{u} \beta_\mathrm{u}} \cos \phi_0 \\
u_0^{'} & = & - \sqrt{\frac{\epsilon_\mathrm{u}}{\beta_\mathrm{u}}} \sin \phi_0 \, .
\label{eq:initcond}
\end{eqnarray}

\begin{figure}[ht]
	\begin{center}
		\includegraphics[width=6cm]{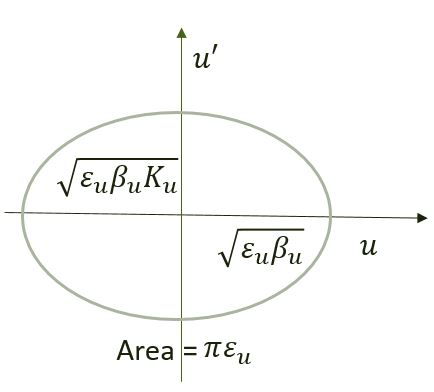}
		\caption{Phase space diagram for motion in a weak-focussing circular accelerator}
		\label{fig:phasespaceweak}
	\end{center}
\end{figure}

\begin{figure}[ht]
	\begin{center}
		\includegraphics[width=6cm]{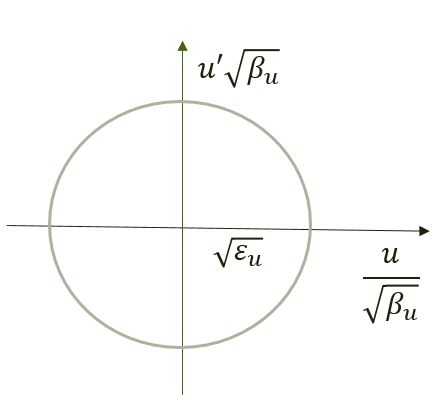}
		\caption{Normalized phase space diagram for motion in a weak-focussing circular accelerator}
		\label{fig:phasespaceweaknorm}
	\end{center}
\end{figure}

The betatron phase in Eq.~(\ref{eq:phase}) grows linearly with the azimuthal position $s$, i.e.,  $\phi_\mathrm{u} (s) = \sqrt{K_\mathrm{u}} s $  and the betatron tunes become $Q_\mathrm{u}  = \frac{\sqrt{K_\mathrm{u}}C_0}{2\pi} $ . Finally, from Eqs.~(\ref{eq:wkfctr2}) and (\ref{eq:wkfctr}), we see that the quantity
 \begin{equation}
 u^2(s) + \frac{{u^{'}(s)}^{2}}{K_\mathrm{u}} = u_0^2 + \frac{{u_0^{'}}^{2}}{K_\mathrm{u}} = \epsilon_\mathrm{u} \beta_\mathrm{u}
 \end{equation}
 is a motion {\em invariant}, i.e., it is independent of $s$.  The form of this equation suggests a convenient visualization of the motion in a {\em phase space} diagram (Fig. (\ref{fig:phasespaceweak}))  in which the axes are position and angle and the particle trajectory is an ellipse. The motion invariant above is related to the area of this ellipse by:
 \begin{equation}
 Area = \pi \epsilon_\mathrm{u}
 \end{equation}
 In practical applications, it is often useful to consider a {\em normalized} phase space representation in which the variables $\frac{u}{\sqrt{\beta_\mathrm{u}}}$ and $\sqrt{\beta_\mathrm{u}} u^{'}$ are used instead. In this representation, particle trajectories are simply circles (Fig. (\ref{fig:phasespaceweaknorm})) and motion over a section of the accelerator (from $s=s_0$ to $s=s_1$ ) is simply a clockwise rotation by an angle equal to the betatron phase advance $\phi_\mathrm{u} = \int_{s_0}^{s_1} \frac{ds}{\beta_\mathrm{u}(s)}$ between $s_0$ and $s_1$

\begin{figure}[ht]
	\begin{center}
		\includegraphics[width=6cm]{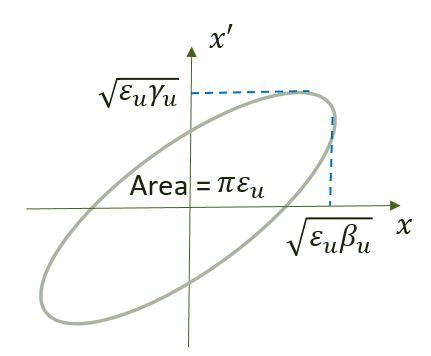}
		\caption{Phase space diagram for motion in a strong-focussing circular accelerator}
		\label{fig:phasespacestrong}
	\end{center}
\end{figure}

We have now obtained a full description of particle motion for the simple case of weak focussing: it is simple harmonic motion with different characteristic frequencies or tunes in each plane (horizontal/vertical). We have determined the conditions for stable motion, calculated the betatron functions and corresponding betatron phase advances, and determined an invariant of the motion. The fact that the betatron functions are constant means that the transverse oscillation frequencies are also constant along the whole accelerator, i.e. the restoring forces that keep electrons stably oscillating around the design orbit are constant.
In the more general case of strong focussing, for which the assumption of azimuthal symmetry is no longer valid, the linear motion can still be described by Eq.~(\ref{eq:pseudosol}) but the betatron functions are no longer constant and the phase advances no longer grows at a steady pace along the accelerator. Instead, in those parts of the machine where the betatron functions are small, the phase advances quickly, i.e., particles oscillate at large frequencies, and conversely particles oscillate more slowly where the betatron functions are larger. One can however still find an invariant of the motion (the Courant--Snyder invariant):

\begin{equation}
\epsilon_\mathrm{u} = \gamma_\mathrm{u} \left(s \right) u^2\left( s \right) + 2 \alpha_\mathrm{u} (s) u \left( s \right) u^{'}\left( s \right) + \beta_\mathrm{u} (s) {u^{'} (s)}^2 \, ,
\end{equation}
where $\alpha_\mathrm{u}(s)$ and $\gamma_\mathrm{u}(s)$ are defined as
\begin{eqnarray}
\alpha_\mathrm{u}(s) & = & -\frac{1}{2}\frac{d\beta_\mathrm{u}(s)} {ds} \\
\gamma_\mathrm{u}(s) & = & \frac{1+\alpha_\mathrm{u}(s)^2}{\beta_\mathrm{u}(s)} \, .
\end{eqnarray}
The motion can be visualized in phase space as well (Fig.~\ref{fig:phasespacestrong}) and the trajectories are still ellipses (as could be inferred from the form of the Courant--Snyder invariant), but now the ellipse is rotated: the rotation angle depends on the Twiss parameter $\alpha_\mathrm{u}(s)$ and varies along the accelerator. The area of the ellipse is, however, still constant and given by $A=\pi\epsilon_0$ just as in the weak-focussing case.  A normalized phase space with variables ($\frac{u}{\sqrt{\beta_\mathrm{u}}}$ ,$\frac{\alpha_\mathrm{u} + \beta_\mathrm{u} u^{'}}{\sqrt{\beta_\mathrm{u}}}$) again turns particle trajectories into circles and reduces all complexity of the motion along the full accelerator to a simple rotation, just as in the weak-focussing case.

\begin{figure}[ht]
	\begin{center}
		\includegraphics[width=6cm]{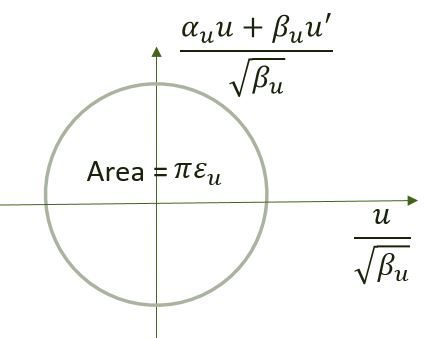}
		\caption{Normalized phase space diagram for motion in a strong-focussing circular accelerator}
		\label{fig:phasespacstrongnorm}
	\end{center}
\end{figure}

All that was said above is valid for an on-energy particle, i.e., one that has exactly the design energy. For particles that show a small energy or momentum deviation, the equation of motion (\ref{eq:eqmotion1}) for the horizontal plane has an additional (non-homogeneous) term . Using know theorems of the theory of linear differential equations, we can write the general solution of Eq.~(\ref{eq:eqmotion1}) as a linear superposition of a general solution to Eq.~(\ref{eq:hill}) and a particular solution to Eq.~(\ref{eq:eqmotion1}). A particular solution of Eq.~(\ref{eq:eqmotion1}) for $\frac{\Delta p}{p_0}=1$ which has the additional property of being periodic with period $C_0$ is called the dispersion function $\eta(s)$, which satisfies
\begin{equation}
\eta^{''}(s) + \left[ \frac{1}{\rho (s)^2} - K(s) \right] \eta (s) = \frac{1}{\rho (s)} \, .
\label{eq:dispersion}
\end{equation}

\begin{figure}[ht]
	\begin{center}
		\includegraphics[width=12cm]{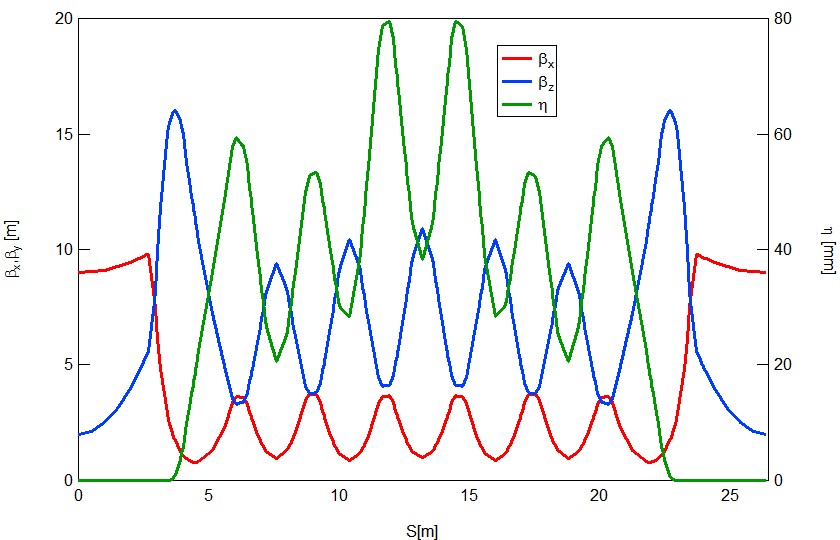}
		\caption{Twiss parameters for one superperiod of the MAX IV 3 GeV storage ring}
		\label{fig:twissMAXIV}
	\end{center}
\end{figure}

The functions $\beta_\mathrm{u}(s)$, along with their derived quantities $\alpha_\mathrm{u}(s)$ and $\gamma_\mathrm{u}(s)$ (collectively known as {\em Twiss parameters}) and the dispersion function $\eta (s)$ are global properties of the ring, defined by how the focussing is distributed along the accelerator, and do not depend on initial conditions of any specific particle in the beam. They provide us with a powerful tool to describe linear motion, in that they allow us to  write the detailed evolution of position and angle for {\em any} particle given its initial conditions. As an example, Fig. \ref{fig:twissMAXIV} shows the betatron and dispersion function calculated for the MAX IV 3 GeV ring.

\subsection{Perturbations in transverse beam dynamics}
In an ideal world of perfect magnets and mechanical structures, the equations of motion discussed in the previous section and their solutions contain all we need to know (in a linear approximation) to design a circular accelerator in a way that guarantees motion stability over many turns. In real life however, magnets and their mechanical support structures as well as the power supplies that provide current to their coils will always suffer from a number of inevitable imperfections. As a result, the actual distribution of bending $\rho(s) $ and focussing $K_\mathrm{u}(s)$ strengths along the accelerator will always differ by a (hopefully) small amount from their  nominal values.  Moreover, the real world will also include a number of non-linear components to the fields distribution, which are not taken into account by the first order treatment we have seen so far. It is therefore of paramount importance that the accelerator design is also robust against such small perturbations.

Consider first the case of linear perturbations -- either in the bending or focussing strengths. The impact of such perturbations may be described by the expressions:
\begin{eqnarray}
\Delta{}u_\mathrm{{c.o.}}(s) & = & \frac{\sqrt{\beta_\mathrm{u}(s){\beta{}}_\mathrm{u}(0)}{\theta{}}_0 \cos⁡(\phi_\mathrm{u} \left(s \right) - \pi{} Q_\mathrm{u})} {2 sin⁡(\pi{} Q_\mathrm{u})} \\
\frac{\Delta{}\beta_\mathrm{u}(s)}{\beta_\mathrm{u}(s)} & =& \frac{{\beta_{}}_\mathrm{u}(0)}{2\sin{\left( 2\pi{}Q_\mathrm{u}\right)}} \cos⁡( 2\pi \phi _\mathrm{u}(s) - 2\pi{}Q_\mathrm{u})\Delta{} K_\mathrm{u}L \, .
\end{eqnarray}

The first equation gives the change in closed orbit (i.e., the periodic solution to the equations of motion) due to a bending strength perturbation localized at $s=0$ and quantified by the {\em kick} angle $\theta$ . The second equation gives the relative change in betatron function resulting from a focussing strength error localized at $s=0$ and characterized by the integrated strength error $\Delta{} K_\mathrm{u}L$. Two important features can be immediately seen form these equations: first, the impact is larger when the perturbations are located at positions where the beta functions are large (proportional to $\sqrt{\beta_\mathrm{u}}$ for the bending strength perturbation and proportional to $\beta_\mathrm{u}$ for the focussing strength perturbation). Second, we note the strong dependence with the betatron tune $Q_\mathrm{u}$: as $Q_\mathrm{u}$ approaches an integer (for the bending strength perturbation) or a half-integer (for the focussing strength perturbation), both expressions diverge!

\begin{figure}[ht]
	\begin{center}
		\includegraphics[width=8cm]{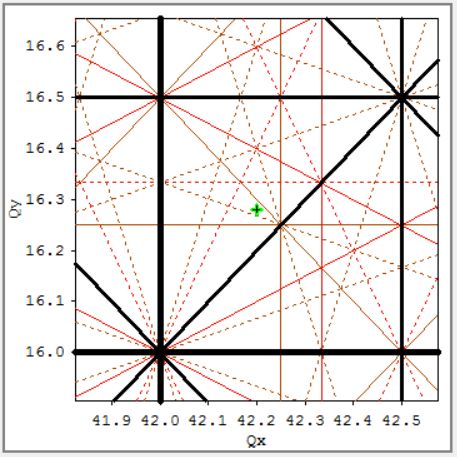}
		\caption{Resonance diagram for the MAX IV 3 GeV ring. The cross indicates the chosen operating point of the machine.}
		\label{fig:resonance}
	\end{center}
\end{figure}

This is actually, a behaviour one might expect from any oscillatory system, that is perturbed or driven at a frequency close to its natural oscillating frequencies, i.e., the system responds resonantly. For example, a bending error perturbation at a fixed position in the accelerator is seen by the passing beam as a periodic kick and, if the phase of the oscillatory motion at every passage is incorrectly chosen, the effects of this sequence of kicks will accumulate from turn to turn leading to unstable motion. In fact, in a storage ring, any combination of horizontal and vertical tunes satisfying $mQ_\mathrm{x} + nQ_\mathrm{z} = p$ where $m,n,p$ are integers (see Fig.~\ref{fig:resonance}) could potentially lead to instability if the appropriate perturbations are in place. Fortunately, not all of these resonances are equally damaging as there are always damping phenomena that help keep the beam stable, so that typically one needs to worry only about the lowest order resonances.

Before leaving the subject of perturbations to the transverse motion, we need to consider also non-linear perturbations. In order to understand why, let us consider what happens to a particle that has an energy slightly different from nominal. We have already seen (cf. Eq.~(\ref{eq:dispersion})) that such a particle has also a slightly different closed orbit that is given by the dispersion function $\eta$

\begin{equation}
x_\mathrm{c.o.}(s)=\frac{\Delta p}{p_0} \eta (s)\, .
\end{equation}

But that is not all: clearly a particle with an energy higher than nominal will experience smaller focussing strengths and as a result, also the betatron functions, phase advances, and tunes will change. The global effect of an energy deviation on the particle tunes is called the chromaticity defined by

\begin{equation}
\xi _\mathrm{u} = \frac{\Delta Q_\mathrm{u}}{\frac{\Delta p}{p_0}} \, .
\end{equation}

Without any correction, chromaticity is a negative number since higher energies lead to less focussing and therefore lower betatron oscillation frequencies. Even if this is in principle not a problem from a single-particle dynamics perspective,  the chromaticity needs to be correct in order to avoid coherent instabilities driven by the fields that are generated by the beam itself. This phenomenon, commonly referred to as the {\em head--tail effect}, could severely limit the maximum attainable current. Chromaticity correction can be accomplished by adding elements to the accelerator magnet lattice which produce a focussing strength that is dependent on the transverse position at which particles cross them and placing those elements at locations in the accelerator (dispersive sections) where there is a correlation between a particle's energy and its transverse position.  These elements are non-linear or sextuple lenses and they render the linear approximation to calculating particle motion inadequate. In fact, while  linear oscillations can remain stable irrespective of the oscillation amplitude, the stability of non-linear oscillations may be restricted to a limited range of amplitudes characterized by the accelerator's {\em dynamic aperture}, which can severely compromise the overall accelerator performance through reduced beam lifetime and injection efficiency.
The proper optimization of such non-linear elements and the correction of the aberrations caused by them is a research topic and a huge amount of effort to find  optimal solutions has been made over the past several decades.

\subsection{Longitudinal dynamics}
\label{sect:longdyn}
So far, we have tacitly assumed that the energy of the particles is constant, even if it may be slight different from the nominal energy. However, as electrons radiate they lose energy and these losses need to be compensated by a radio-frequency system. At one or several locations around the ring, radio-frequency cavities  store energy in the form of stationary electromagnetic waves with a longitudinal component of the electric field so that at each passage through the cavity, electrons can get a longitudinal {\em kick} that restores the lost energy. Since the fields in the cavity oscillate at the frequency $f_\mathrm{rf}$, only electrons that cross the cavity at the right moment to recover exactly the amount of energy they lost in one turn will remain at a constant energy. The question then arises: what happens to electrons that have a small deviation in time of arrival at the cavities and therefore get the wrong energy kick? And what about an electron that has a small deviation in energy and gets just the longitudinal kick that was supposed to be given to the particle with nominal energy? Fortunately, there is an underlying self-stabilizing mechanism at play that guarantees that, as long as the oscillation amplitudes are not too large, the motion remains stable and the particle's energy and time of arrival at the cavities  merely oscillate around the nominal values instead of wandering off to ever larger deviations that would eventually lead to particle loss. This is called the principle of phase stability and can be described like this: particles that have energies slightly larger than nominal will have a larger radius of curvature in the bending magnets and therefore go through a longer path around the ring, causing them to arrive later at the cavities, since the speed is almost independent of energy for highly relativistic particles. So, as long as particles that arrive late at the cavities receive less energy than nominal, their deviations with respect to the nominal particle for the next turn will be smaller. The inverse will be true for particles with less energy than nominal so that the net results is that deviating particles are always brought back towards the nominal particle. All we need is to make sure particles cross the cavity at a time when the accelerating fields are on a negative (decreasing) slope. Of course that solution has limitations and only works up to a certain energy or time deviation amplitude. Indeed, since the fields vary sinusoidally, the {\em restoring force} of this oscillating system is intrinsically non-linear, defining maximum energy and phase apertures. Still, just like for the transverse motion, at small amplitudes, the motion looks like a harmonic oscillator with a characteristic frequency $f_\mathrm{s}$ called the synchrotron frequency, which, when expressed in units of the revolution frequency $f_0$ gives the synchrotron tune $Q_\mathrm{s}$.

\subsection{Equilibrium beam parameters}
As we have seen above, and electron beam in a storage ring can be seen as a collection of many three-dimensional oscillators. If the accelerator design parameters are properly chosen, the oscillations in the transverse (betatron) and longitudinal (synchrotron) planes can all be made stable within a certain range of oscillation amplitudes.  As long as the beam intensity is low, these oscillators are all uncoupled, but as the current grows, fields produced by the electrons themselves act back on the beam coupling their motion.

In an electron storage ring, the interplay between the average effect of the emission of synchrotron radiation  (which acts as a damping mechanism, a sort of friction that leads to a reduction of oscillation amplitude in all planes) and the quantized nature of the photon emission process (which acts like a heating mechanism exciting betatron and synchrotron motion, leading to an increase of oscillation amplitudes) lead to the establishment of an equilibrium distribution of particles in all three planes of motion characterized by the equilibrium horizontal and vertical beam emittances $\varepsilon_\mathrm{u}$ and equilibrium relative energy spread $\sigma _{\delta \mathrm{p}}$. These quantities are functions of the detailed lattice design and, together with the Twiss parameters, may be used to determine the equilibrium beam dimensions at any azimuthal position $s$ by

\begin{eqnarray}
{\sigma{}}_\mathrm{x}\left(s\right) & = & \sqrt{{\epsilon{}}_\mathrm{x}{\beta{}}_x\left(s\right)+{{\sigma{}}_{\delta{}}}^2{\eta{}\left(s\right)}^2}
\\
{\sigma{}}_\mathrm{y}\left(s\right) & = & \sqrt{{\epsilon{}}_\mathrm{y}{\beta{}}_\mathrm{y}\left(s\right)}\, .
\end{eqnarray}

\section{How the accelerator vacuum system affects accelerator performance: a few illustrations}

\begin{figure}[ht]
	\begin{center}
		\includegraphics[width=12cm]{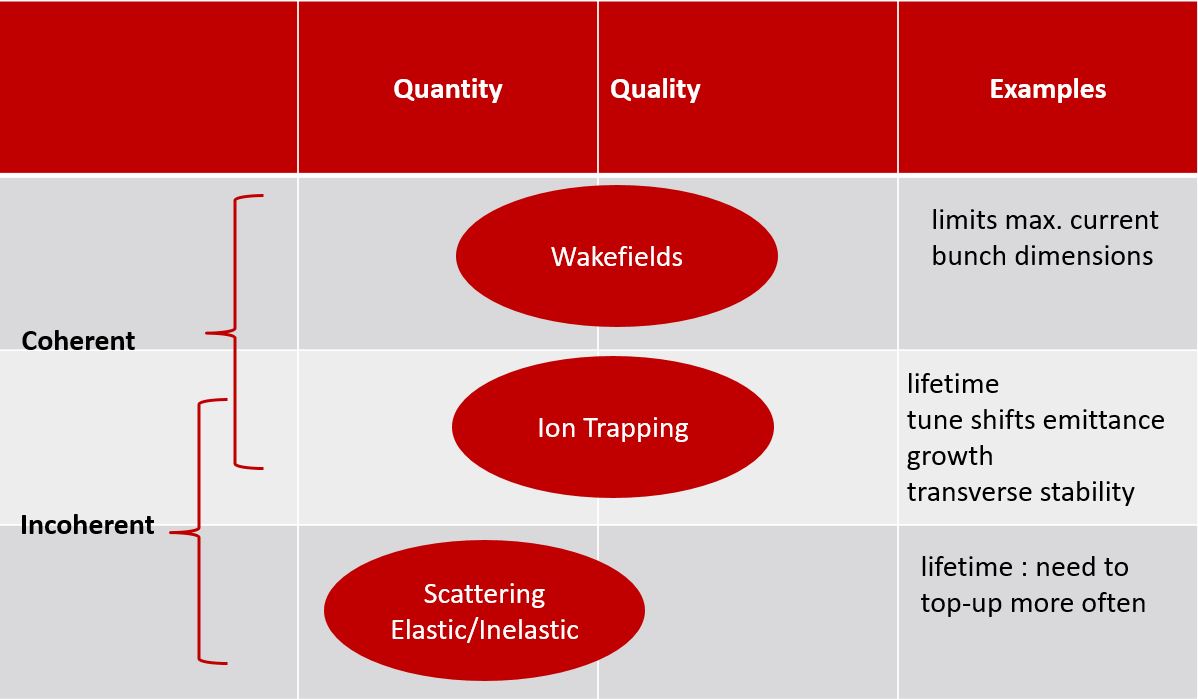}
		\caption{How the vacuum system affects accelerator performance}
		\label{fig:vacsys}
	\end{center}
\end{figure}

The vacuum system design is crucial to the overall accelerator performance in a number of ways (Fig. \ref{fig:vacsys}) setting limits to the intensity (number of circulating electrons) or the quality of the beam (beam size, beam stability) or to the rate at which beam particles are lost (i.e., the beam lifetime). These effects can be broadly classified as incoherent or coherent, depending on whether they are associated with individual encounters of electrons with residual gas molecules (scattering) or if they are related to motion of the beam as a whole, for example due to the excitation of wake fields by the beam that are modified by the walls of the vacuum chamber and act back on the beam. Additionally, collisions of circulating electrons with residual gas molecules produce positive ions that can be captured (trapped)  by the beam leading to a reduction of beam lifetime due to the increased local gas pressure, betatron tune shifts and spreads due to the partial beam neutralization generated by the ion cloud, and eventually emittance growth and collective instabilities. Such effects had serious consequences in some early machines, but typically become less severe as the vacuum pressure improve after sufficient vacuum conditioning.

Let us consider some of the incoherent effects that affect the beam lifetime, as an illustration of how the concepts introduced earlier in this lecture are relevant in understanding how to best keep these phenomena under control. Take for example, elastic scattering of beam electrons by the nuclei of residual gas molecules. The basic physics is given by the Rutherford scattering cross-section:
\begin{equation}
\frac{d{\sigma{}}_\mathrm{el}}{d\Omega{}}={\left(\frac{1}{4\pi{}{\epsilon{}}_0}\frac{Z{e_0}^2}{2p_0c}\right)}^2\frac{1}{{\sin{\left(\frac{\theta{}}{2}\right)}}^4} \, ,
\end{equation}
where $\theta$ is the scattering angle and $Z$ the atomic number of the target nucleus. Particles will be lost in such a scattering event if the scattering angle is larger than
\begin{equation}
{\theta{}}_\mathrm{max}^2=\frac{{\left(\frac{A_\mathrm{u}^2}{\beta_\mathrm{u}{}}\right)}_\mathrm{min}}{{\beta{}}_\mathrm{u0}} \, ,
\end{equation}

where $A_\mathrm{u}$ is the physical chamber aperture at the location $s$ and the scattering is assumed to happen at location $s=0$. The normalized rate of particle loss (i.e., the inverse lifetime associated to these collisions) is then
\begin{equation}
\frac{1}{{\tau{}}_\mathrm{el}}=-\frac{1}{N}\frac{dN}{dt}=2\pi{}cn\int_{{\theta{}}_\mathrm{max}}^{\pi{}}\frac{d{\sigma{}}_\mathrm{el}}{d\Omega{}}\sin{\theta{}}d\theta{} \, ,
\end{equation}
where $n$ is the volume density of residual gas nuclei. These results indicate that elastic scattering becomes more important as the beam energy and apertures goes down. One should also be careful to avoid high-$Z$ gases and high vacuum pressures specially at the locations of large beta functions.

As a second example, let us consider inelastic scattering on nuclei, i.e., events in which a collision of a high energy electron with the nucleus of a residual gas molecules results in emission of photons through the process of bremsstrahlung. As the photon is emitted, the electron lose energy and if the sudden energy change is larger than the maximum accepted energy deviation by the RF \query{AQ: please define RF} system, the particle will be lost\footnote{Here I assume the RF system determines this limitation, as described in section \ref{sect:longdyn}, but in practice this limit is often set by non-linear phenomena associated with the magnet lattice design.}.

The probability of losing an energy $\delta$  in such a collision event is given by the cross-section:
\begin{equation}
\frac{d{\sigma{}}_\mathrm{BS}}{d\delta{}}=\frac{{\alpha{}}_f4Z^2r_0^2}{\delta{}}\left\{\left[\frac{4}{3}\left(1-\frac{\delta{}}{E}\right)+{\left(\frac{\delta{}}{E}\right)}^2\right]\ln{\left(\frac{183}{Z^{\frac{1}{3}}}\right)}+\frac{1}{9}\left(1-\frac{\delta{}}{E}\right)\right\} \, ,
\end{equation}
and the corresponding lifetime, for an assumed energy acceptance  $\delta _\mathrm{acc}$ becomes

\begin{equation}
\begin{split}
\frac{1}{{\tau{}}_\mathrm{BS}} & = -\frac{1}{N}\frac{dN}{dt}=
cn\int_{{\delta{}}_\mathrm{acc}}^E\frac{d{\sigma{}}_\mathrm{BS}}{d\delta{}}d\delta{}
\\
 & = cn4{\alpha{}}_\mathrm{f}Z^2r_e^2\left\{\frac{4}{3}\left(ln\left(\frac{E}{{\delta{}}_\mathrm{acc}}\right)-\frac{5}{8}\right)\ln{\left(\frac{183}{Z^{\frac{1}{3}}}\right)}+\frac{1}{9}\left(\ln{\left(\frac{E}{{\delta{}}_{acc}}\right)}-1\right)\right\} \, ,
\end{split}
\end{equation}

where $\alpha_\mathrm{f}$ is the fine structure constant and $r_0$ the classical electron radius.

Again, high-$Z$ gases are the ones to worry about, whereas the dependence on the beam energy and energy acceptance is rather weak.

\end{document}